\documentclass[runningheads]{llncs}
\usepackage{graphicx}
\usepackage{cite}
\usepackage{amsmath,amssymb,amsfonts}
\usepackage{algorithmic}
\usepackage{graphicx}
\usepackage{textcomp}
\usepackage{xcolor}
\usepackage{graphicx, tabto}
\usepackage{amssymb}
\usepackage{array}
\usepackage{multirow}
\usepackage{booktabs}
\usepackage{colortbl}
\usepackage{arydshln,graphicx,xcolor,array}
\usepackage{tikz}
\usetikzlibrary{matrix}
\usepackage{blindtext}
\usepackage[export]{adjustbox}
\usepackage{cite}
\usepackage{hyperref}
\usepackage[font=small,skip=0pt]{caption}
\usepackage{subcaption}
\def\BibTeX{{\rm B\kern-.05em{\sc i\kern-.025em b}\kern-.08em
    T\kern-.1667em\lower.7ex\hbox{E}\kern-.125emX}}
\usepackage{comment}

\begin{document}
\title{\large{Brain Image Segmentation of 2D Images using U-Net}}
%


% \author{Submission ID: 7}
% \institute{Main Manuscript}

\author{
Angad R.S. Bajwa\inst{1,2}
}
%
% \\ TC is supported by the EPSRC SeeBiByte grant and the UKRI DART programme.
\authorrunning{A.R.S. Bajwa}
\institute{Indian Institute of Technology (IIT), Varanasi (BHU) \and
National Institute of Technology (NIT), Tiruchirappalli, India.}
\maketitle              % typeset the header of the contribution
\begin{abstract}
Brain tumour segmentation is an essential task in medical image processing. Early diagnosis of brain tumours plays a crucial role in improving treatment possibilities and increases the survival rate of the patients. Manual segmentation of the brain tumours for cancer diagnosis, from large number of MRI images, is both a difficult and time-consuming task. There is a need for automatic brain tumour image segmentation. The purpose of this project is to provide an automatic brain tumour segmentation method of MRI images to help locate the tumour accurately and quickly.

\end{abstract}
\section{Introduction}
Tumors may start in the brain, or cancer elsewhere in the body may spread to the brain. Four standard MRI modalities used in brain image study are  T1-weighted MRI (T1), T2-weighted MRI (T2), T1-weighted MRI with gadolinium contrast enhancement (T1-Gd), and Fluid Attenuated Inversion Recovery (FLAIR).Generally, T1 images are used for distinguishing healthy tissues, and in this project, we use T1 brain images for our study.

Often automatic brain tumor segmentation methods use hand-crafted features such as edges, corners, histogram of gradient, local binary pattern. These features are extracted and then given to the classifier. The training procedure of the classifier is not affected by the nature of those features. 
Semantic segmentation is commonly used in medical imaging to identify the precise location and shape of structures in the body and is essential to the proper assessment of medical disorders and their treatment. 
Recently, deep convolutional neural networks (CNNs) have led to drastic improvements in numerous computer vision tasks.\cite{ref_modal}

In this project, we try to predict the brain tumor location using one such CNN architecture- U Net. Open source code for the project is available (link in Acknowledgement section). We study three types of U-Net architectures. One is the simple U-Net, the second is a U-Net with additional skip connections, and lastly, an M- Net, a new unique approach to semantic segmentation. All previous works on this dataset are for the classification of tumor types—none of the previous works performed on this dataset are intended for segmentation purposes. 

\begin{figure*}[h]
\centering
\includegraphics[width=\textwidth]{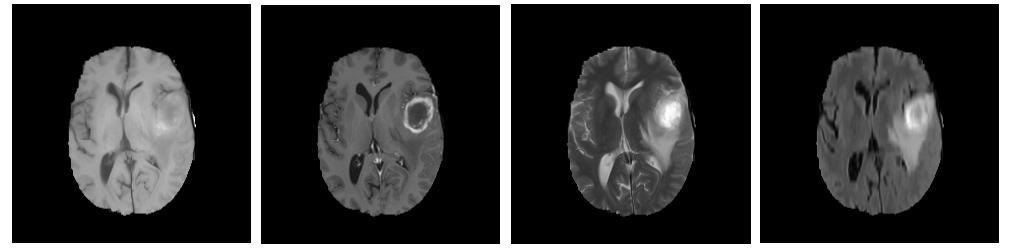}
\caption{Four different MRI modalities showing a high grade glioma, each enhancing
different subregions of the tumor. From left; T1, T1-Gd, T2, and FLAIR \cite{ref_review}} \label{fig1}
\end{figure*}

\section{Dataset and Evaluation Metrics}

This brain tumor T1-Weighted MRI image-dataset consists of 3064 slices. There are 1047 coronal images. Coronal images are those which are captured from the back of the head. Axial images, those taken from above the skull, are 990 in number. The dataset also contains 1027 sagittal images that are captured from the side of the skull. This dataset has a label for each image, identifying the type of the tumor. These 3064 images belong to 233 patients. The size of each image is 512*512 pixels. The dataset includes three types of tumors- 708 Meningiomas, 1426 Gliomas, and 930 Pituitary tumors, which are publicly available: \href{ http://dx.doi.org/10.6084/m9.figshare.1512427}{\color{blue} here}

We evaluate the segmentation results using the Dice coefficient. The Dice coefficient, also called the overlap index, is a metric used for validation in medical image segmentation. The pair-wise overlap of the segmentations is calculated by:

\begin{equation}
    \centering
    Dice   coefficient = \frac{2 \cdot TP}{2 \cdot TP+FP+FN}
\end{equation}

where TP is true positive results or correctly segmented tumour pixels, FP is false positive, and FN is the false negative results of the segmentation. False positive results are obtained when a pixel which is non-tumorous is classified as tumorous. Also, FN refer to the number of pixels that are tumorous and are falsely labelled as non-tumorous.\cite{ref_bayesian}

\begin{equation}
    \centering
    Dice  loss = 1.0 - Dice coefficient
\end{equation}

\begin{equation}
    \centering
    BCE loss = \frac{-1}{output size}\sum\limits_{i=1}^{output size}{y_{i}log(y^{*}_{i})+ (1-y_{i})log(1-y^{*}_{i})}
\end{equation}

Where y is the target value (in our case 1 or 0, tumor or non-tumor pixel) and y* is the predicted value for that pixel. Output size is the image size in our 512*512.
We use a loss function which is a sum of both the binary cross entropy and dice loss.
\begin{figure*}[h]
\centering
\includegraphics[width=\textwidth]{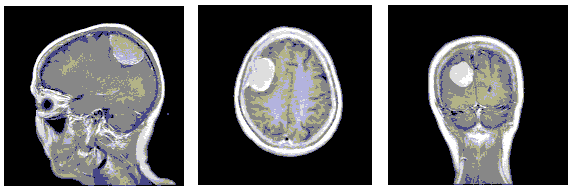}
\caption{Brain MRI slices captured from different directions \cite{ref_type} } \label{fig1}
\end{figure*}

\section{Deep Architectures}

\begin{figure*}[h]
\centering
\includegraphics[width=\textwidth]{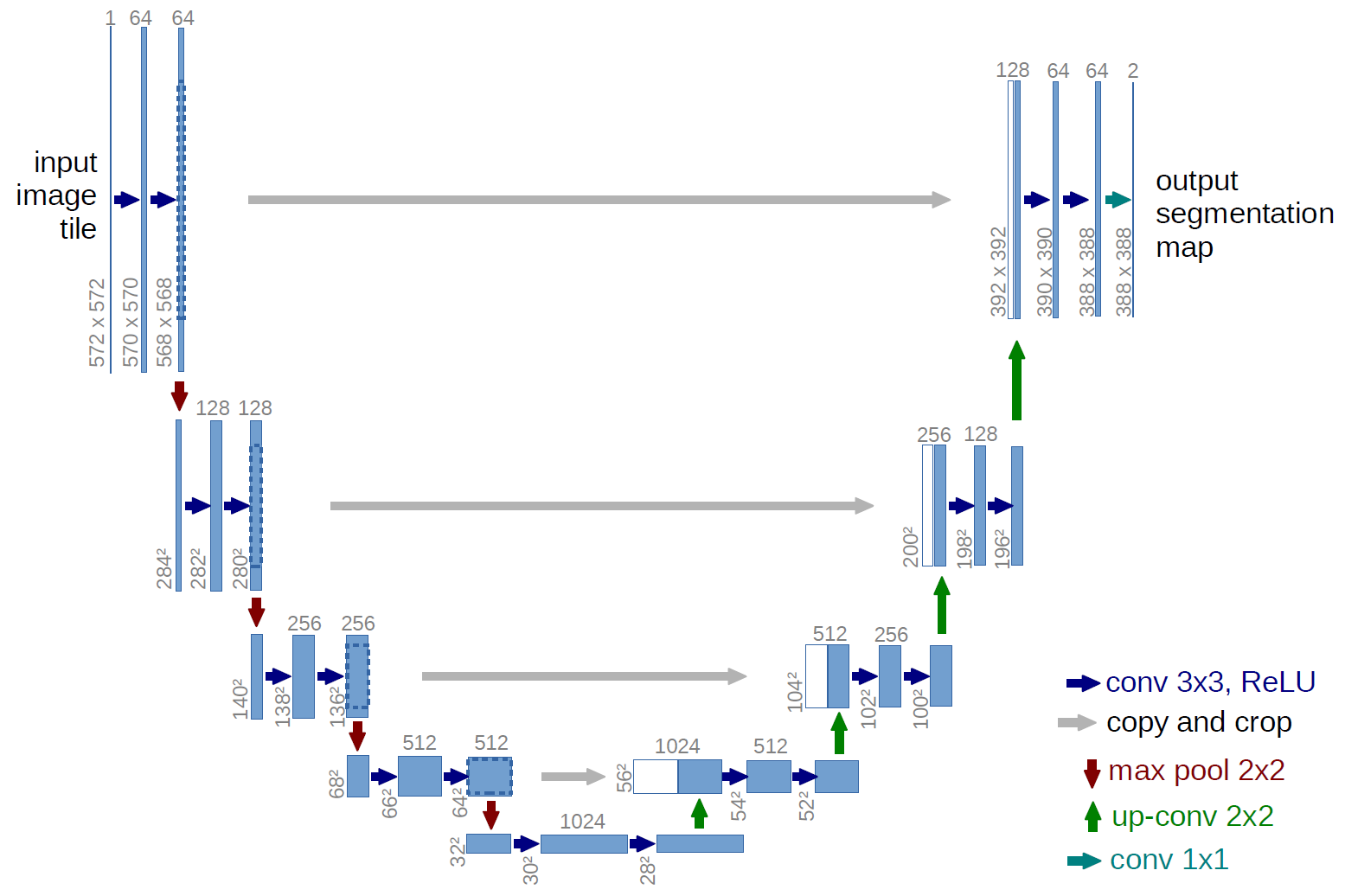}
\caption{U-Net architecture \cite{ref_unet}} \label{fig1}
\end{figure*}

\subsubsection{\textbf{U-Net:}}
\label{section:U-Net}

U-net architecture was introduced in the paper” U-Net: Convolutional Networks for Bio-medical Image Segmentation” in March 2015. It was developed by Olaf Ronneberger for Bio-Medical Image Segmentation.\cite{ref_unet} The U-net architecture contains two paths. The first path is the contraction path (encoder part) which is used to capture the context of the image. The encoder is a stack of convolutional and max pooling layers. The second path is an expanding path (decoder part) to enable precise localization using transposed convolutions. It is an end to end  fully convolutional network which only contains convolutional layers and no Dense layers because of which such architecture cannot accept images of any size. Two consecutive convolutions layers are applied before max-pooling. This is done as when pooling is applied, lot of information is lost as the dimensions of the data are halved. So, convolution layers are stacked before each pooling, so it can build up
better representations of the data without quickly losing all the spatial information.\cite{ref_temporal}

\subsubsection{\textbf{U Net with skip(both short and long) connections:}}
\label{section:U-Net}
To the U-Net architecture I have added skip  connections between the first convolution layer and the max pooling layer on each level. The main purpose of this was to increase accuracy and uniform distribution of parameters in the layers. \\

% \textbf{Training} 

% This served as the baseline model for our upcoming methods. After experimenting with different network architectures, one with $5$ convolutional blocks followed by  Global Average Pooling\cite{ref_gap} and a classification layer was chosen for this purpose. Each convolution block is further, a combination of one convolutional layer followed by ReLU activation, dropout, Max Pooling and batch normalization layers. We used convolutional kernels with spatial extent $7,5,3,3,3$ for consecutive convolutional blocks having $16,32,64,128$ and $256$ feature maps respectively. The final classification layer contains $7$ (\textit{number of classes}) neurons and SoftMax is used as the activation function to obtain a probability distribution at the output.

% \subsection{CNN based Transfer learning:}
% \label{sectin:TL}
% Transfer learning\cite{ref_tl} is an effective technique for bypassing the problem of limited data availability in different domains of learning algorithms. In this approach, knowledge gained by a model for solving a particular task is utilized for relevant tasks. Several state of the art models including VGG-16\cite{ref_vgg}, ResNet50\cite{ref_res}, Efficient Nets\cite{ref_eff} pre-trained on Imagenet \cite{ref_imagenet} dataset were tested and based on a moderate performance measure EfficientNet B0 is selected for this purpose. Last few layers of the model are modified according to our problem and fine-tuned entirely before using for inference purpose.

\subsubsection{\textbf{M-Net:}}
\label{section:global_att} M-net mainly has four pathways of 2D filters: two main encoding and decoding paths and two side paths, which give our architecture deep supervision. Each pathway has four steps. In the encoding path, each step has a cascade of 2D convolution filters of size 3x3 and maxpooling by 2x2, reducing the input size by half and allowing the network to learn contextual information. In the cascade of convolution filters, skip connection is introduced to enable the network to learn better features. The decoding layer is identical to encoding layers with one exception: maxpooling is replaced by an upsampling layer to double the input size and recover an output image of the original size. Similarly, skip connections are also implemented between corresponding encoding and decoding layers to ensure that the network has sufficient information to derive fine grain labeling of an image without post-processing. The left leg operates on 256*256 with four maxpooling layers of size 2x2, and the outputs are given as input to the corresponding encoding layers. The right leg upsamples the output of each of the decoding layers to the original size of 256*256. Finally, the decoding layer and the right leg output is processed by a 1x1 convolution layer with L channels, where L is the number of structures of interest, including background. 
The advantage of the M-net is that barring one 3D convolution filter, all other filters are 2D filters which allows end-to-end training of the network with considerably low memory requirement (~5GB).\cite{ref_mnet}

\begin{figure*}[h]
\centering
\includegraphics[width=\textwidth]{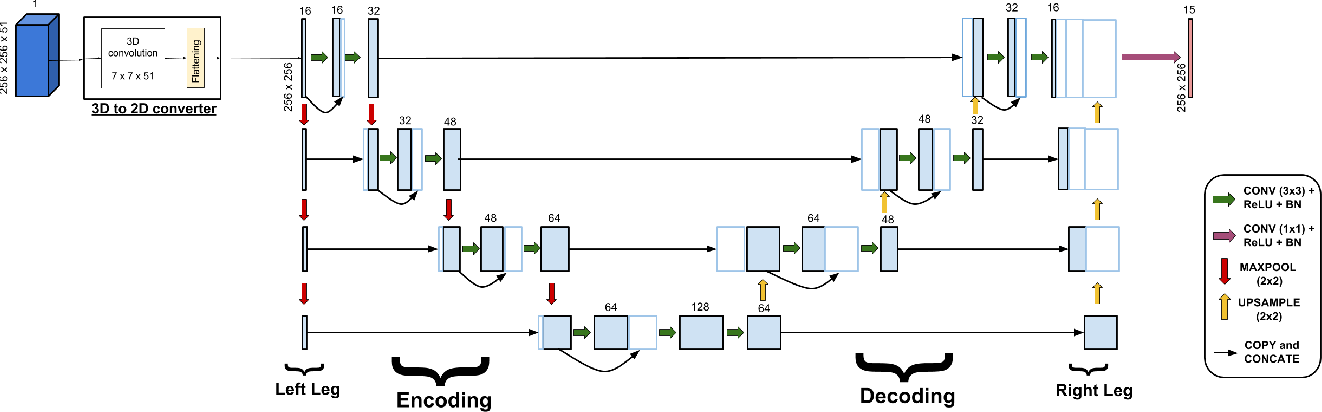}
\caption{M-Net architecture \cite{ref_mnet}} \label{fig1}
\end{figure*}

\begin{table*}[h!]
\centering
\caption{Performance measure of models}
\begin{tabular}{|c|c|c|c|}
\hline
\begin{tabular}[c]{@{}c@{}} Model \end{tabular} & Epochs & Loss & Dice Score \\ \hline
U-Net                                &    50  &     0.0577    & 67.28           \\ \hline
U-Net with additional skip connection                                                                     &    50     &      0.0537   &  61.02    \\
\hline
M-Net 
 &  50   &   0.0765    &  64.55    
\\ \hline
\end{tabular}
\label{tab:model performance}
\end{table*}

\subsection{Training}
\label{section:training}
We split the data into  training, validation and test sets in the ratio of 80:10:10. Thereby having 2451 training images, 307 validation images and 306 test images. We implement our method using Keras with a Tensorflow backend using ADAM optimizer (learning rate=0.0001). The model is trained for 50 epochs.

Due to limited RAM and GPU we split the training into 2 compartments. The first with 30 epochs, saving the model and then training further for the next 20 epochs.Once the model is trained, we evaluate it on the test set.

\section{Results:}
We use Google Colab to implement the model. Colab uses Intel(R) Xeon(R) CPU @ 2.20GHz, 12 GB of RAM and NVIDIA Tesla K80 TPU.

On comparing the three models, we find that the simple U- Net performs the best with a dice score of 67.2.

\begin{figure*}[]
\centering
\includegraphics[width=0.75\textwidth]{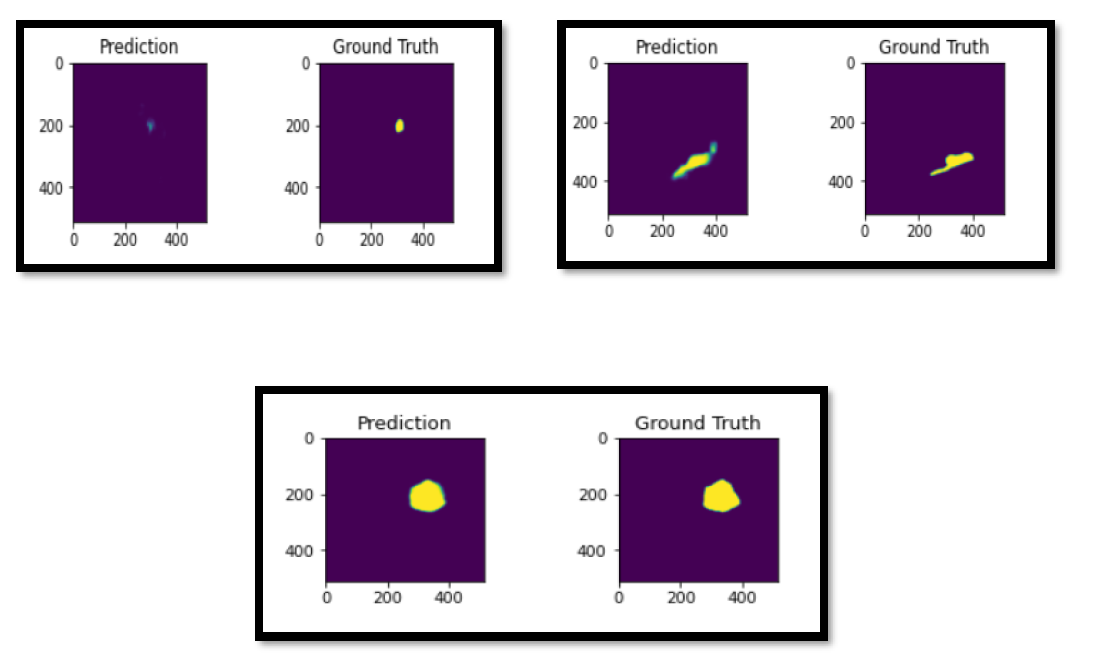}
\caption{U-Net: prediction vs ground truth for unseen data} \label{fig1}
\end{figure*}

\section{Conclusion}

We experimented with three different U-Net architectures and tried two loss functions to derive the best results. The simple U- Net performs the best on the given dataset. We also introduced skip connections between convolution filters and deep supervision functionality in our network, which allows it to learn better features. We have come up with a memory-efficient way of detecting tumors that can reduce both the time and difficulty faced in clinics. Early discovery can prevent patients from entering severe stages of the tumor. 

\begin{figure*}[]
\centering
\includegraphics[width=0.75\textwidth]{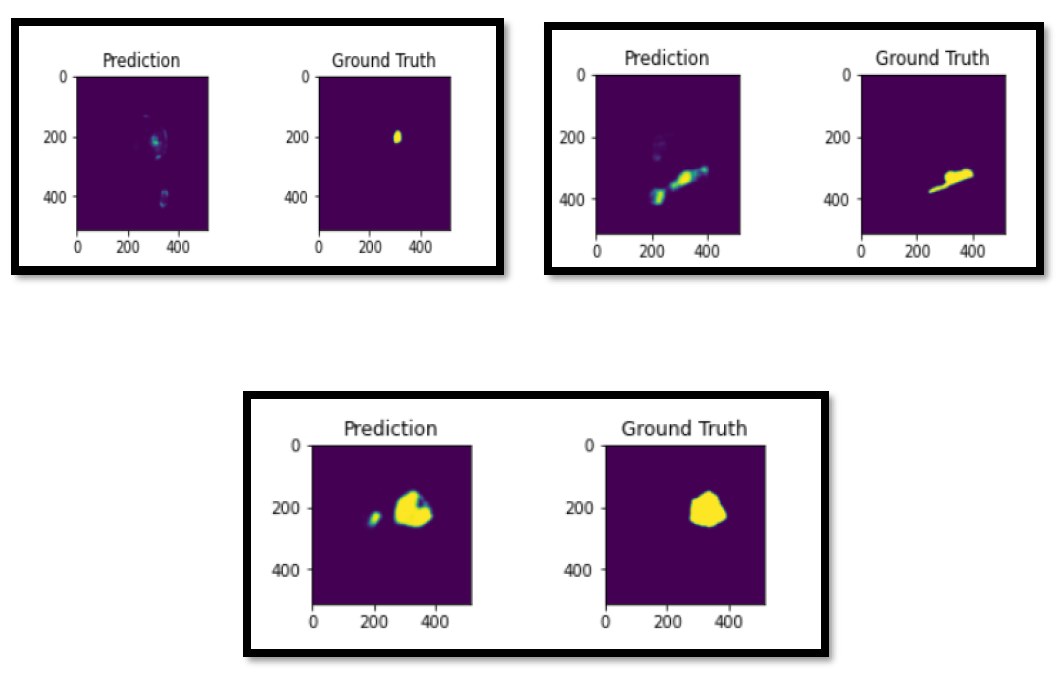}
\caption{U-Net + additional skip connection: prediction vs ground truth for unseen data} \label{fig1}
\end{figure*}

\begin{figure*}[]
\centering
\includegraphics[width=0.75\textwidth]{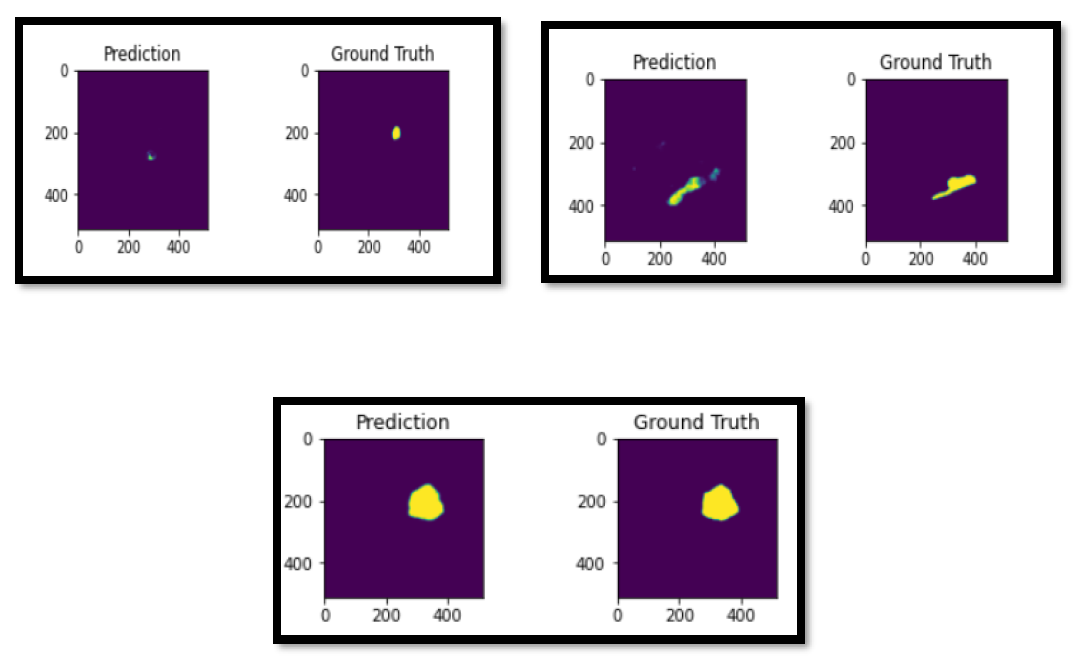}
\caption{M-Net: prediction vs round truth for unseen data} \label{fig1}
\end{figure*}

\section{Acknowledgement}

This work was done under the supervision of Professor Rajeev Srivastava during my research internship with his group at Indian Institute of Technology (Varanasi) in 2020. I am grateful for the training and mentoring I received from Prof Srivastava during the project. Open source code for the project is available at https://github.com/angadbajwa23/Segmentation-of-2D-Brain-MR-Images-using-Deep-Neural-Architectures.

\end{document}